\documentclass[twocolumn]{aa}
\usepackage{graphicx}
\usepackage{txfonts}
\begin{document}
%-------------------------------------------------------------------------------
\title{Supernova 2000cb: high-energy version of SN~1987A}

\author{V. P. Utrobin\inst{1,2} \and N. N. Chugai\inst{3}}

\institute{
   Max-Planck-Institut f\"ur Astrophysik,
   Karl-Schwarzschild-Str. 1, 85741 Garching, Germany
\and
   Institute of Theoretical and Experimental Physics,
   B.~Cheremushkinskaya St. 25, 117218 Moscow, Russia
\and
   Institute of Astronomy of Russian Academy of Sciences,
   Pyatnitskaya St. 48, 119017 Moscow, Russia}

\date{Received 26 April 2011 / Accepted 6 July 2011}

%===============================================================================
\abstract{% Context.
Among type IIP supernovae there are a few events that resemble
   the well-studied supernova 1987A produced by the blue supergiant in the
   Large Magellanic Cloud.
}{% Aims.
We study a peculiar supernova 2000cb and compare it with the supernova 1987A.
}{% Methods.
We carried out hydrodynamic simulations of the supernova in an extended
   parameter space to describe its light curve and spectroscopic data.
The hydrogen H$\alpha$ and H$\beta$ lines are modeled using a time-dependent
   approach.
}{% Results.
We constructed the hydrodynamic model by fitting the photometric and
   spectroscopic observations.
We infer a presupernova
   radius of $35\pm14 R_{\sun}$, an ejecta mass of $22.3\pm1 M_{\sun}$,
   an explosion energy of $(4.4\pm0.3)\times10^{51}$ erg, and a radioactive
   $^{56}$Ni mass of $0.083\pm0.039 M_{\sun}$.
The estimated progenitor mass on the main sequence lies in the range of
   $24-28 M_{\sun}$.
The early H$\alpha$ profile on Day 7 is consistent with the density
   distribution found from hydrodynamic modeling, while the H$\alpha$ line
   on Day 40 indicates an extended $^{56}$Ni mixing up to a velocity of
   8400 km\,s$^{-1}$.
We emphasize that the dome-like light curves of both supernova 2000cb and
   supernova 1987A are entirely powered by radioactive decay.
This is unlike normal
   type IIP supernovae, the plateau of which is dominated by the internal energy
   deposited after the shock wave propagation through the presupernova.
We find signatures of the explosion asymmetry in the photospheric and
   nebular spectra.
}{% Conclusions.
The explosion energy of supernova 2000cb is higher by a factor of three
   compared to supernova 1987A, which poses a serious problem for explosion
   mechanisms of type IIP supernovae.
}
%-------------------------------------------------------------------------------
%
\titlerunning{Energetic SN~2000cb}
\authorrunning{V. P. Utrobin and N. N. Chugai}
\maketitle

%===============================================================================
\section{Introduction}
\label{sec:intro}
%-------------------------------------------------------------------------------
Type II-plateau supernovae (SNe IIP) represent the most numerous subclass of
   core-collapse SNe.
They are characterized by a $\sim100$ day plateau in the light curve, which is
   a generic feature of the explosion of a red supergiant (RSG) star
   (Grassberg et al. \cite{GIN_71}; Falk \& Arnett \cite{FA_77}).
This picture is in line with the theory of stellar evolution that predicts
   that stars in the range of $9-25...30~M_{\sun}$ end their life as RSGs
   (Heger et al. \cite{HFWLH_03}).
The RSG nature of type IIP pre-SNe was confirmed by the detection
   of pre-SNe in archival images (Smartt \cite{Sma_09}).

SN~1987A in the Large Magellanic Cloud (LMC), which was identified with
   the explosion of a blue supergiant (BSG), became a serious challenge for
   theoreticians.
There is still no clear answer to why the pre-SN was a small-radius star.
Two major explanations were proposed:
   (i) the mixing between the He-core and the hydrogen envelope mediated by
   a fast rotation in combination with the low metallicity
   (Saio et al. \cite{SNK_88}; Weiss et al. \cite{WHT_88};
   Woosley et al. \cite{WPE_88});
   (ii) the loss of an extended hydrogen envelope in a close binary system
   (Hillebrandt \& Meyer \cite{HM_89}; Podsiadlowski \& Joss \cite{PJ_89}).
The unresolved issue of the pre-SN origin in the case of SN~1987A is a strong
   stimulus for studying objects that resemble one another.
Pastorello et al. (\cite{PBB_05}) studied a peculiar SN~1998A,
   another example of the SN originating in the explosion of a BSG.
This SN is very similar photometrically and spectroscopically to SN~1987A,
   although its explosion energy probably was somewhat higher than that of
   SN~1987A (Pastorello et al. \cite{PBB_05}).

Recently, Kleiser et al. (\cite{KPK_11}) have analyzed photometric
   and spectroscopic observations of SN~2000cb and conclude that the SN was
   related with the explosion of a BSG.
In that sense SN~2000cb is another counterpart of SN~1987A.
Nevertheless, authors have emphasized differences, particularly, in the light curve shape
   and in expansion velocities, which were substantially higher in SN~2000cb.
Using scaling relations and SN~1987A as a template, Kleiser et al. (\cite{KPK_11})
   have derived the ejecta mass of $\sim16.5~M_{\sun}$ and the explosion energy of
   $\sim4\times10^{51}$ erg for SN~2000cb.
On the other hand, hydrodynamic modeling has led Kleiser et al. (\cite{KPK_11})
   to the modest estimate of $\sim1.7\times10^{51}$ erg, comparable to that of
   SN~1987A.
It should be noted, however, that authors did not attain a reasonable fit of
   the bolometric light curve.
The question of reliable SN~2000cb parameters therefore remains open.

Motivated by the importance of the SN~1987A-like events and rather
   complete observational light curve of SN~2000cb (Kleiser et al. \cite{KPK_11}),
   we revisit the analysis of this SN by employing a standard
   hydrodynamic model (Utrobin \cite{Utr_04}, \cite{Utr_07}).
A brief description of the hydrodynamic model is presented in Sect.~\ref{sec:hydmod}.
Because the SN parameters cannot be recovered from the photometric data alone,
   one also needs to use the line profiles in which the mass-velocity distribution
   is imprinted.
By this approach we obtain the basic parameters of the optimal model given in
   Sect.~\ref{sec:result-param}, while in Sect.~\ref{sec:result-asym} we address
   signatures of the explosion asymmetry in the photospheric and
   nebular spectra.
In Sect.~\ref{sec:disc} we discuss results and their implications for
   the explosion mechanism problem, and in Sect.~\ref{sec:concl} we summarize
   the conclusions.

Following Kleiser et al. (\cite{KPK_11}) we adopt a distance to the spiral galaxy
   IC 1158 of 30 Mpc, a reddening $E(B-V)=0.114$ mag, an explosion date of
   April 21.5 UT (JD 2\,451\,656), and a recession velocity to the host galaxy of
   1918.67 km\,s$^{-1}$.

%===============================================================================
\section{Model overview}
\label{sec:hydmod}
%-------------------------------------------------------------------------------
%
\begin{figure}[t]
   \resizebox{\hsize}{!}{\includegraphics[clip, trim=0 0 0 20]{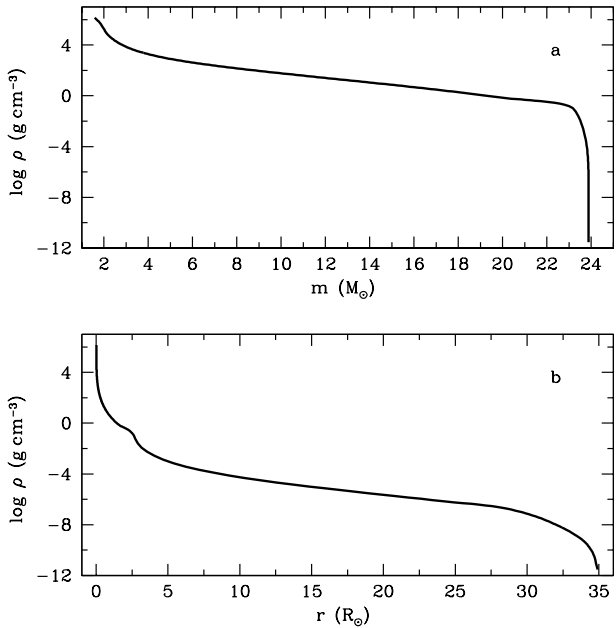}}
   \caption{%
   Density distribution as a function of interior mass \textbf{a}) and
   radius \textbf{b}) for the optimal pre-SN model of SN~2000cb.
   The central core of 1.6 $M_{\sun}$ is omitted.
   }
   \label{fig:denmr}
\end{figure}
\begin{figure}[b]
   \resizebox{\hsize}{!}{\includegraphics[clip, trim=0 0 0 240]{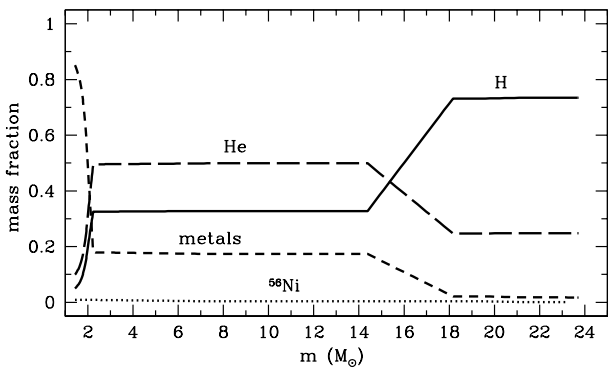}}
   \caption{%
   The mass fraction of hydrogen (\emph{solid line\/}), helium
      (\emph{long dashed line\/}), heavy elements (\emph{short dashed line\/}),
      and radioactive $^{56}$Ni (\emph{dotted line\/}) in the ejecta of
      the optimal model.
   }
   \label{fig:chcom}
\end{figure}
The hydrodynamic code with one-group radiation transfer (Utrobin \cite{Utr_04},
   \cite{Utr_07}) is used to simulate the SN~2000cb event.
The explosion is modeled by placing the supersonic piston close to the outer
   edge of the 1.6 $M_{\sun}$ central core, which is removed from the
   computational mass domain and assumed to collapse to become a neutron star.
The hydrodynamic model in spherically-symmetric approximation is determined by
   the explosion energy ($E$), the ejecta mass ($M_{env}$), the pre-SN radius
   ($R_0$), and the amount of ejected $^{56}$Ni ($M_{\mathrm{Ni}}$).
The last can be reliably determined at the radioactive tail from
   the bolometric luminosity or from the flux in broad-band filter by adopting
   SN~1987A as a template.
Parameters $E$, $M_{env}$, and $R_0$ are determined primarily by the plateau
   luminosity, the plateau duration, and the velocity of the line-absorbing gas
   at the photospheric epoch.
The velocity constraint is usually provided by the absorption minima of weak
   lines.
However, the minima are poorly defined in the case of SN~2000cb which is
   characterized by broad and blended lines.
We therefore rely on the H$\alpha$ and H$\beta$ line profiles, which are studied
   with the atmosphere model in the time-dependent approach (Utrobin \& Chugai
   \cite{UC_05}).

The atmosphere model is based on the time-dependent ionization and excitation
   kinetics of hydrogen and other elements, the time-dependent kinetics of
   molecular hydrogen, and the time-dependent energy balance for the gas
   temperature (Utrobin \& Chugai \cite{UC_05}).
The density distribution, chemical composition, radius of the photosphere,
   and effective temperature are provided by the hydrodynamic model.
The continuum radiation escaping the photosphere is set to be a black body one
   with the effective temperature and the photospheric brightness corresponding
   to the one obtained by our models for SN~1987A.
The obtained time-dependent structure of the atmosphere is then used to
   model synthetic spectra at selected epochs.
The spectra are simulated by means of the Monte Carlo technique.
The Sobolev local escape approximation is assumed for the line radiation
   transfer dominated by the line absorption.
The line emissivity and the Sobolev optical depth are determined by level
   populations, which in turn are provided with the time-dependent approach.
The Thomson scattering on free electrons, Rayleigh scattering on neutral
   hydrogen, and the relativistic effects are also taken into account.
Photons striking the photosphere are assumed to be diffusively reflected back
   into the atmosphere with the albedo calculated according to Chugai \& Utrobin
   (\cite{CU_00}).

It is well established that the evolutionary pre-SN is not the best choice for
   modeling the SN~IIP event (Utrobin \& Chugai \cite{UC_08}, \cite{UC_09}).
This is also true for SN~2000cb (cf. Kleiser et al. \cite{KPK_11}).
We, therefore, prefer to search for the appropriate pre-SN model
   among nonevolutionary hydrostatic RSG configurations that retain major
   features of the evolutionary models.
The essential property of the nonevolutionary model is a smoothed jump in
   density and chemical composition between the helium core and the hydrogen
   envelope.
The adopted pre-SN model of SN~2000cb is shown in Fig.~\ref{fig:denmr}, while
   the distribution of major constituents is displayed in Fig.~\ref{fig:chcom}.
As we see below, the ejecta mass of SN~2000cb is slightly higher than SN~1987A.
We therefore assume the mass of the unmixed helium-core to be $9~M_{\sun}$,
   which corresponds to the final helium core of a main-sequence star of
   $\approx25~M_{\sun}$ (Hirschi et al. \cite{HMM_04}).
It is noteworthy that the result is not sensitive to the helium-core mass,
   because the mixing effects dominate the variation in the helium-core
   mass (Utrobin \cite{Utr_07}).
We adopt the ejected metal core material (mostly oxygen) to be $3~M_{\sun}$,
   which is close to the amount ejected by a $25~M_{\sun}$ main-sequence star
   (Woosley \& Weaver \cite{WW_95}).
These metals are assumed to be homogeneously mixed in the helium core, and their
   mixing with the hydrogen envelope follows the mixing of helium
   (Fig.~\ref{fig:chcom}).
The $^{56}$Ni mixing is another important ingredient of the model, because it markedly
   affects the light curve shape as well.
Moreover, as we will see, the $^{56}$Ni mixing can also be effectively probed
   by the hydrogen line profiles.

%===============================================================================
\section{Results}
\label{sec:result}
%-------------------------------------------------------------------------------
%===============================================================================
\subsection{Basic parameters}
\label{sec:result-param}
%-------------------------------------------------------------------------------
%
\begin{figure}[t]
   \resizebox{\hsize}{!}{\includegraphics[clip, trim=0 0 0 120]{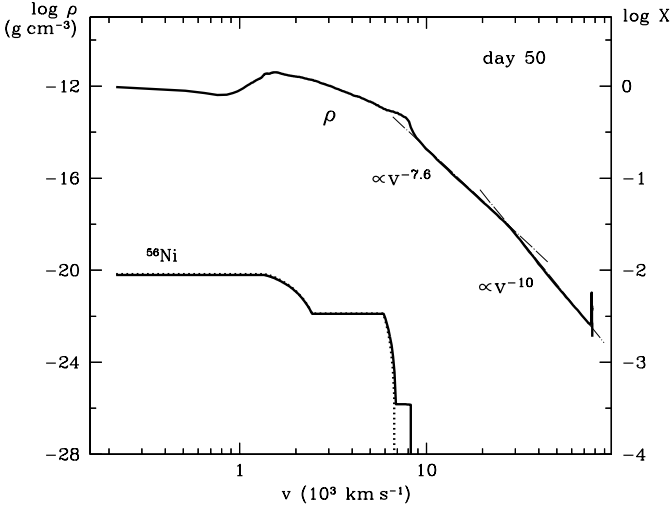}}
   \caption{%
   The density and the $^{56}$Ni mass fraction as a function of velocity
      for the optimal model at $t=50$ days (\emph{solid lines\/}).
   \emph{Dash-dotted} straight lines are the power-law fits of density.
   The $^{56}$Ni distribution is shown for the moderate mixing in the range
      $v\leq 6800$ km\,s$^{-1}$ (\emph{dotted line\/}) and the strong mixing,
      $v\leq 8400$ km\,s$^{-1}$, in the optimal model (\emph{solid line\/}).
   }
   \label{fig:deni}
\end{figure}
The SN~2000cb parameters are found by computing a grid of the hydrodynamical
   models and the H$\alpha$ line.
The optimal model simultaneously fits the bolometric light curve and
   the H$\alpha$ line profile fairly well for
   the ejecta mass $M_{env}=22.3\pm1~M_{\sun}$,
   the explosion energy $E=(4.4\pm0.3)\times10^{51}$ erg,
   the pre-SN radius $R_0=35\pm14~R_{\sun}$, and
   the $^{56}$Ni mass $M_{\mathrm{Ni}}=0.083\pm0.039~M_{\sun}$.
The uncertainties in basic parameters are calculated by assuming a 23\% error
   in the distance and a 5\% error in the characteristic duration of the light
   curve according to Kleiser et al. (\cite{KPK_11}) and by estimating a 5\% error
   in the photospheric velocity.
It is interesting that the pre-SN radius of $35~R_{\sun}$ is equal to that of
   SN~1987A (Utrobin \cite{Utr_05}).
Remarkably, the amount of $^{56}$Ni found from the bolometric light curve
   coincides with the estimate we obtain using the $I$ magnitude at
   the radioactive tail and SN~1987A as a template.
This means that the empirical bolometric luminosity recovered from the filtered
   fluxes is reliable.

\begin{figure}[t]
   \resizebox{\hsize}{!}{\includegraphics[clip, trim=0 0 0 240]{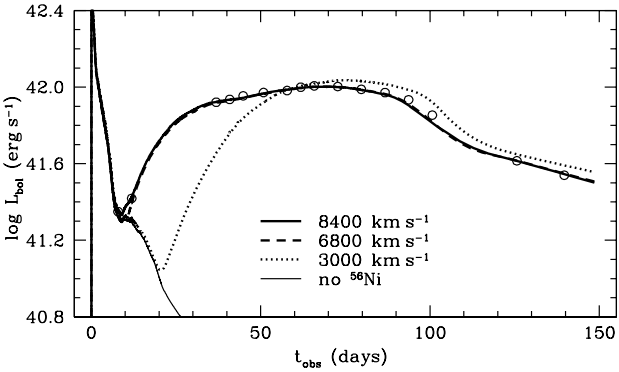}}
   \caption{%
   The bolometric light curve of the optimal model (\emph{thick solid line\/})
      overplotted on the empirical bolometric light curve of SN 2000cb recovered
      from the $BVI$ magnitudes reported by Kleiser et al. (\cite{KPK_11})
      (\emph{open circles\/}).
   \emph{Dashed line} represents the model with the moderate $^{56}$Ni mixing
      to $6800$ km\,s$^{-1}$ compared to the one in the optimal model up to
      $8400$ km\,s$^{-1}$ (Fig.~\ref{fig:deni}).
   \emph{Dotted line} shows the light curve of the model with the $^{56}$Ni
      mixing to $3000$ km\,s$^{-1}$ observed in SN~1987A.
   The light curve in the absence of radioactive $^{56}$Ni is shown by a
      \emph{thin solid line}.
   }
   \label{fig:blc}
\end{figure}
The density distribution of the freely expanding envelope (Fig.~\ref{fig:deni})
   is nearly flat in the velocity range $v\lesssim4000$ km\,s$^{-1}$ and drops in the
   outer layers with the steepening increasing outward.
This behavior can be
   described by a broken power law with an effective index of $-7.6$ and $-10$.
The density spike in the outermost layers (Fig.~\ref{fig:deni}) is a shell with
   a mass of $\sim8\times10^{-7}~M_{\sun}$ formed by the shock breakout
   (Grassberg et al. \cite{GIN_71}; Chevalier \cite{Che_81}).
The boundary velocity is $\approx76\,000$ km\,s$^{-1}$, substantially higher
   than the boundary velocity of SN~1987A, $\approx36\,000$ km\,s$^{-1}$
   (Utrobin \cite{Utr_04}).

The model bolometric light curve for the case without radioactive $^{56}$Ni
   shows that the initial luminosity peak during the first ten days is produced
   by the release of the internal energy deposited after the shock wave
   propagation through the pre-SN (Fig.~\ref{fig:blc}).
The peak is substantially broader and more luminous than both the
   observed and model peaks of SN~1987A, and this reflects the greater explosion
   energy of SN~2000cb.
Remarkably, the major broad maximum of the light curve with a time scale of
   $\sim100$ days, which resembles a wide dome in shape, is entirely powered
   by the $^{56}$Ni$\,\rightarrow^{56}$Co$\,\rightarrow^{56}$Fe decay
   (Fig.~\ref{fig:blc}), as is the case for SN~1987A.
The physics of the dome-like light maximum turns out to be essentially the same as
   in SNe~Ia and SNe~Ib/c.
The different physics of the light curve on a time scale of $100$ days in the
   normal SNe~IIP and the SN~1987A-like events is crucial for understanding
   the origin of a distinction between these categories of SNe~II.

\begin{figure}[t]
   \resizebox{\hsize}{!}{\includegraphics[clip, trim=0 0 0 120]{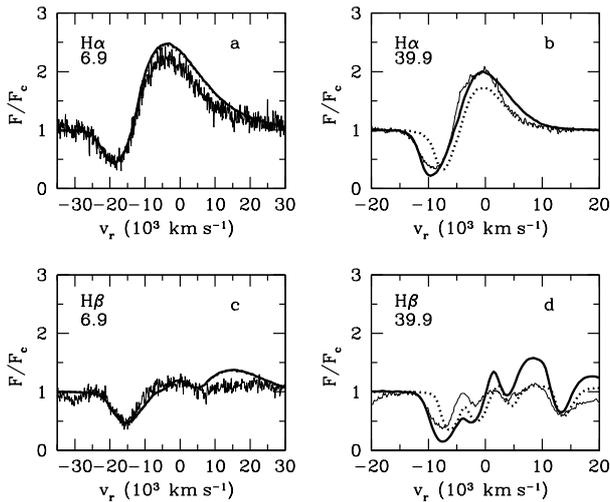}}
   \caption{%
   The H$\alpha$ and H$\beta$ lines on Days 7 and 40.
   The model profiles are calculated for two cases of $^{56}$Ni mixing,
      strong (\emph{thick solid line\/}) and moderate (\emph{dotted line\/}),
      and overplotted on the observed profiles (\emph{thin solid line\/}).
   Modeling the H$\beta$ profile includes the Fe\,II 4924, 5018,
      5169 \AA\, and Ba\,II 4935 \AA\ lines.
   }
   \label{fig:ha}
\end{figure}

An amount of the $^{56}$Ni mixing is probed by the rising part of the light curve
   after Day 10 (Fig.~\ref{fig:blc}).
It is evident that the $^{56}$Ni mixing to $3000$ km\,s$^{-1}$ observed in
   SN~1987A is inconsistent with the observed light curve, so it indicates
   stronger mixing.
Indeed, the light curves for both versions of the extended (to $6800$ km\,s$^{-1}$
   and $8400$ km\,s$^{-1}$) $^{56}$Ni mixing (Fig.~\ref{fig:deni}) agree well
   with observations (Fig.~\ref{fig:blc}).
A step-like feature of the $^{56}$N distribution in the range of $6800-8400$
   km\,s$^{-1}$ (Fig.~\ref{fig:deni}), which contains $10^{-3}~M_{\sun}$ of
   $^{56}$Ni, is needed to account for the H$\alpha$ and H$\beta$ absorptions
   at high velocities on Day 40 (Figs.~\ref{fig:ha}b and d).
Remarkably, the strong H$\beta$ absorption on Day 40 is in sharp contrast
   to the almost undetectable H$\beta$ line in the spectrum of SN~1987A at the same
   epoch (Hanuschik \cite{HTS_89}).
This reflects the different situation with the $^{56}$Ni mixing in SN~1987A
   where the radioactive $^{56}$Ni is mixed to substantially lower velocity
   ($\approx3000$ km\,s$^{-1}$).

It should be emphasized that the early H$\alpha$ and H$\beta$ line profiles
   on Day 7 are not sensitive to the $^{56}$Ni distribution at all
   (Figs.~\ref{fig:ha}a and c), while on Day 40 the absorption is very
   much sensitive to the $^{56}$Ni distribution (Figs.~\ref{fig:ha}b and d).
The contributions of the density and $^{56}$Ni distributions in the H$\alpha$ and
   H$\beta$ absorptions thus turn out to be decoupled.
This permits us to use the early H$\alpha$ absorption to probe the density
   in the outer layers and the late time H$\alpha$ absorption to probe
   an amount of $^{56}$Ni mixing in the outer layers.
We therefore conclude that the high velocity estimated from the blue absorption
   wing of the H$\alpha$ and H$\beta$ profiles on Day 7 can serve as
   an independent indicator of the small pre-SN radius and the high explosion
   energy of SN~2000cb.

Generally, hydrogen in the outer layers might be alternatively excited by
   X-rays from the ejecta/wind interaction.
We calculated the X-ray luminosity from the forward and reverse shocks,
   assuming the mass-loss rate $\dot{M}=10^{-6}~M_{\sun}$ yr\,$^{-1}$ and
   the wind velocity $u=500$ km\,s$^{-1}$, i.e., adopting the wind as ten times
   denser than around SN~1987A (Chevalier \& Dwarkadas \cite{CD_95}).
Even in this optimistic case, the X-ray luminosity on Day 40 is
   $\sim10^{35}$ erg\,s$^{-1}$, which is able to provide a nonthermal
   excitation rate of hydrogen not greater than $10^{-4}$ of what is produced
   by $^{56}$Ni.
We therefore rule out any sizable contribution of the ejecta/wind interaction
   in the formation of the high-velocity H$\alpha$ and H$\beta$ absorptions
   on Day 40.

%===============================================================================
\subsection{Explosion asymmetry}
\label{sec:result-asym}
%-------------------------------------------------------------------------------
%
\begin{figure}[t]
   \resizebox{\hsize}{!}{\includegraphics[clip, trim=0 220 0 120]{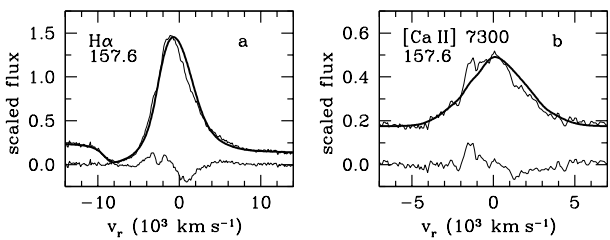}}
   \caption{%
   The H$\alpha$ line and the [Ca\,II] 7300 \AA\ doublet on Day 158.
   The calculated profiles (\emph{thick line\/}) are overplotted on the observed
       profiles (\emph{thin line\/}).
   At the bottom, residuals between the observed profile and the one computed
       for the spherically-symmetric model are shown.
   }
   \label{fig:asym}
\end{figure}
Even though the spherically-symmetric model is consistent with the light curve and
   the early hydrogen line absorptions, on Day 40 the model overproduces
   the flux in the red part of the H$\alpha$ emission component
   (Fig.~\ref{fig:ha}b).
We checked different artificial, spherically-symmetric models and found that
   they all overproduced the flux in the red part of the H$\alpha$ emission
   in the same way.
Moreover, the similar model of the H$\alpha$ line in the normal type IIP
   SN~1999em on Days 26 and 52 does not show any deficit in the red wing
   (Utrobin \cite{Utr_07}).
We therefore suggest that the deficit in the emission in the red part of
   the H$\alpha$ emission is related to the asymmetry of the $^{56}$Ni
   distribution; namely, the front side of the envelope presumably
   contains more $^{56}$Ni than does the rear side of the envelope.
In this picture, the nonthermal excitation in the rear hemisphere turns out to be
   lower, and consequently, the H$\alpha$ emission is weaker than in the front
   side, which thus explains the observed profile asymmetry.

To check whether the $^{56}$Ni asymmetry in SN~2000cb also extends into the
   deeper layers, as in the case of SN~1987A, we have analyzed the
   H$\alpha$ line and the [Ca\,II] 7292, 7319 \AA\ doublet at the nebular
   stage, on Day 158.
The H$\alpha$ emissivity reflects the deposition of gamma-ray energy, which
   is almost local on Day 158 because of the large optical depth of
   $\approx6$ for gamma-ray absorption.
The [Ca\,II] 7300 \AA\ emission may reflect both the asymmetry of the $^{56}$Ni
   distribution and the distribution of Ca itself.
The observed lines (Kleiser et al. \cite{KPK_11}), together with the calculated
   profiles for the spherical model, are shown in Fig.~\ref{fig:asym}.
We computed the lines taking the continuum absorption into account, which is
   predicted by the hydrodynamical model at this stage, and the doublet
   structure of the [Ca\,II] 7300 \AA\ line with the finite optical depth
   of [Ca\,II] doublet lines.

Both the observed H$\alpha$ and [Ca\,II] 7300 \AA\ emissions reveal a blueshift
   with respect to the spherically-symmetric model, which is most apparent in
   the residuals shown in Fig.~\ref{fig:asym}.
The $^{56}$Ni asymmetry in the inner layers can be quantified by the blueshift
   of the mean radial velocity in the range $|v_r|<4000$ km\,s$^{-1}$.
For the observed H$\alpha$ line the flux-weighted shift is $-390$ km\,s$^{-1}$,
   while for the spherical model profile, the mean shift is $-145$ km\,s$^{-1}$
   with the resulting blueshift of $-245$ km\,s$^{-1}$.
For the [Ca\,II] 7300 \AA\ doublet the relative blueshift is $-211$ km\,s$^{-1}$.
The found asymmetry is significantly more than the uncertainty in the radial
   velocity of the SN rest frame related to the host galaxy's rotation.
Indeed, for the inclination angle of IC 1158 $i=62^{\circ}$ taken from LEDA and
   the angle of $\sim37^{\circ}$ between the radius-vector of the SN and the minor axis,
   the velocity uncertainty of the rest frame turns out to be
   $\pm70$ km\,s$^{-1}$ assuming the circular velocity of 240 km\,s$^{-1}$.
We conclude, therefore, that the blueshift of the observed H$\alpha$ and
   [Ca\,II] 7300 \AA\ profiles is real, thus suggesting the predominant ejection
   of $^{56}$Ni in the inner layers within the front hemisphere.

%===============================================================================
\section{Discussion}
\label{sec:disc}
%-------------------------------------------------------------------------------
Our major result is that SN~2000cb was the high-energy explosion of the BSG star.
This confirms the general conclusion of Kleiser et al. (\cite{KPK_11}), who
   ``favor a high-energy explosion of a relatively small radius star, most
   probably a BSG".
The ejecta mass of $22.3\pm1~M_{\sun}$, combined with the collapsing core of
   $1.6~M_{\sun}$, gives the total pre-SN mass of $23.9\pm1~M_{\sun}$.
Adopting the same mass loss by pre-SN of $2.4\pm1~M_{\sun}$ as in type IIP
   SN~2004et of the comparable pre-SN mass (Utrobin \& Chugai \cite{UC_09}),
   for SN~2000cb we come to the progenitor mass on the main sequence of
   $26.3\pm2~M_{\sun}$.
The error in mass includes the uncertainties in the ejecta mass and the mass
   loss.
This progenitor mass is about the upper limit of the main-sequence stars
   presumably exploding as SNe~IIP (Heger et al. \cite{HFWLH_03}).

Kleiser et al. (\cite{KPK_11}) have derived two different sets of the ejecta mass
   and the explosion energy for SN~2000cb.
The first one is based upon the analytical scaling relations and the SN~1987A
   parameters.
This approach results in the ejecta mass of $\approx16.5~M_{\sun}$ and the
   explosion energy of $\sim4\times10^{51}$ erg.
Our ejecta mass is higher by 35\%, while the explosion energy is only 10\%
   higher.
This should be considered as reasonable agreement.
On the other hand, the hydrodynamic modeling of the light curve led Kleiser
   et al. (\cite{KPK_11}) to the lower energy of $1.7\times10^{51}$ erg for the
   ejecta mass of $18~M_{\sun}$.
It should be noted, however, that these authors did not attain an exact fit of
   the observational bolometric light curve and did not use the velocity
   constraints.
This means that the hydrodynamical model of Kleiser et al. (\cite{KPK_11})
   cannot be considered optimal, so the comparison with our
   hydrodynamical model cannot be made in a straightforward way.

\begin{table}[t]
\caption[]{Hydrodynamic models of type IIP supernovae.}
\label{tab:sumtab}
\centering
\begin{tabular}{ l c c c c c c }
\hline\hline
\noalign{\smallskip}
 SN & $R_0$ & $M_{env}$ & $E$ & $M_{\mathrm{Ni}}$ 
       & $v_{\mathrm{Ni}}^{max}$ & $v_{\mathrm{H}}^{min}$ \\
       & $(R_{\sun})$ & $(M_{\sun})$ & ($10^{51}$ erg) & $(10^{-2} M_{\sun})$
       & \multicolumn{2}{c}{(km\,s$^{-1}$)}\\
\noalign{\smallskip}
\hline
\noalign{\smallskip}
 1987A &  35  & 18   & 1.5   & 7.65 &  3000 & 600 \\
1999em & 500  & 19   & 1.3   & 3.6  &  660  & 700 \\
2000cb &  35  & 22.3 & 4.4   & 8.3  &  8400 & 440 \\
 2003Z & 229  & 14   & 0.245 & 0.63 &  535  & 360 \\
2004et & 1500 & 22.9 & 2.3   & 6.8  &  1000 & 300 \\
2005cs & 600  & 15.9 & 0.41  & 0.82 &  610  & 300 \\
2009kf & 2000 & 28.1 & 21.5  & 40.0 &  7700 & 410 \\
\noalign{\smallskip}
\hline
\end{tabular}
\end{table}
A summary of the parameters of the best observed SNe~IIP, which were derived by
   a uniform procedure, is compiled in Table~\ref{tab:sumtab}.
The listed parameters are the pre-SN radius, the ejecta mass, the explosion
   energy, the $^{56}$Ni mass, the maximal velocity of $^{56}$Ni mixing
   zone, and the minimal velocity of the hydrogen-rich envelope.
All SNe~IIP are characterized by a deep mixing of hydrogen, indicated by
   the low value of $v_{\mathrm{H}}^{min}$, which is consistent with
   2D simulations (M\"{u}ller et al. \cite{MFA_91};
   Kifonidis et al. \cite{KPSJM_03}, \cite{KPSJM_06}).
In the case of SN~2000cb, the explosion energy is significantly higher, while
   the $^{56}$Ni mixing occurs within the wider mass and velocity ranges
   than for SN~1987A.
The very high degree of the $^{56}$Ni mixing is a prominent feature of SN~2000cb
   that should provide an important observational constraint on the yet poorly
   known explosion mechanism.

\begin{figure}[t]
   \resizebox{\hsize}{!}{\includegraphics[clip, trim=0 0 0 0]{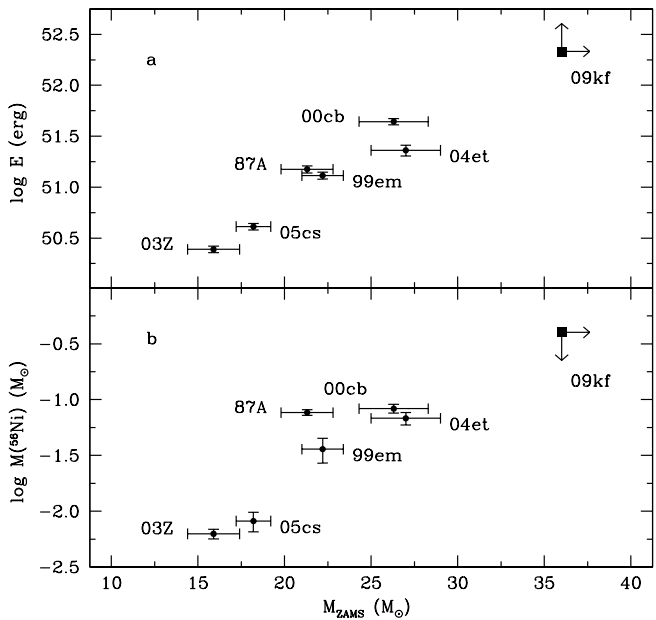}}
   \caption{%
   Explosion energy \textbf{a}) and $^{56}$Ni mass \textbf{b}) versus
      hydrodynamic progenitor mass for seven core-collapse SNe.
   }
   \label{fig:nienms}
\end{figure}
The position of SN~2000cb on the plots of the explosion energy versus
   the progenitor mass (Fig.~\ref{fig:nienms}a) and the total $^{56}$Ni mass
   versus the progenitor mass (Fig.~\ref{fig:nienms}b) is consistent
   with the correlations recovered earlier for the SNe~IIP sample
   studied hydrodynamically in a uniform way (Utrobin et al. \cite{UCB_10}).
The asymmetry of the $^{56}$Ni distribution in SN~2000cb is well established
   by our analysis of the spectra at the photospheric and nebular stages.
The asymmetry of the bulk of radioactive $^{56}$Ni is rather moderate in terms
   of the velocity shift along the line of sight, $\Delta v_z\approx-240$
   km\,s$^{-1}$.
We cannot rule out that the actual deprojected asymmetry is stronger.
This might be revealed by polarization data that are unfortunately missing.
The one-sided ejection of radioactive $^{56}$Ni is not unique feature of
   SN~2000cb, because the line profiles of SN~1987A show evidence that
   the $^{56}$Ni distribution is skewed towards the rear hemisphere
   (Haas et al. \cite{HCE_90}; Chugai \cite{Chu_91}).
The one-sidedness of the $^{56}$Ni ejecta seems to be a common feature for
   SNe~IIP (Chugai \cite{Chu_07}) and indicates that the explosion
   mechanism of both the normal SNe~IIP and the SN~1987A-like events seems
   to be similar.

Two major explosion mechanisms for core-collapse SNe are presently considered
   as promising: neutrino-driven explosion and magneto-rotational mechanism.
Particularly important in this regard is our conclusion about the high value
   of the explosion energy of SN~2000cb.
The neutrino-driven mechanism is currently able to produce the explosion with
   the energy at best $<10^{51}$ erg (Scheck et al. \cite{SJFK_08}), and
   even a number of $10^{51}$ erg seems to be beyond reach of the recent
   3D models (Rantsiou et al. \cite{RBNA_11}).
The energetics of the neutrino-driven explosion mechanism is thus
   at least four times less than what is needed to account for
   the SN~2000cb phenomenon.
This casts serious doubts on the universality of the neutrino-driven mechanism
   for the SN~II explosions.
The current situation with the magneto-rotational mechanism is more
   optimistic: the available 2D simulations provide the explosion energy of
   $(0.5-2.6)\times10^{51}$ erg (Bisnovatyi-Kogan et al. \cite{BMA_08}), by
   a factor of two below the energy of SN~2000cb.

On the other hand, hypernovae with the explosion energy of $(2-3)\times10^{52}$
   erg associated with gamma-ray bursts, e.g. SN~1998bw
   (Iwamoto et al. \cite{IMN_98}) and SN~2003dh (Deng et al. \cite{DTMMN_05}),
   and the type IIP SN2009kf of the similar energy (Utrobin et al. \cite{UCB_10})
   have already posed a problem for the high-energy explosion mechanism.
Presently, we lack firm evidence that a unique explosion
   mechanism operates in SNe~IIP and hypernovae, although the existence of
   phenomena likewise SN~2000cb and SN~2009kf pushes us towards this line of
   reasoning.

The BSG configuration of SN~1987A has been partially attributed to the low
   metallicity of the LMC.
The host galaxy of SN~2000cb is the galaxy IC 1158, normal, although slightly
   subluminous, Sc galaxy with the normal metallicity $12+\log\mbox{[O/H]}=8.7$
   (Anderson et al. \cite{ACJHH_10}).
The deprojected galactocentric distance of the SN is $\approx 8$ kpc, comparable
   to that of the Sun.
It is unlikely, therefore, that the progenitor of SN~2000cb was a low-metallicity
   star, although this cannot be completely precluded.
Mixing favored by the fast rotation or evolution of the close binary, proposed
   for SN~1987A, remains viable possibilities for the pre-SN to become a BSG.
It is noteworthy that overlapping the mass ranges of SN~2000cb and SN~1987A and
   of normal SNe~IIP indicates that the high stellar mass is unlikely a unique
   condition that makes the pre-SN into a BSG.
Even though high mass of the progenitor favors the BSG structure (Woosley et al.
   \cite{WHWL_97}), some additional factor seems to be required to explain
   the compactness of the pre-SN.

A special remark should be made concerning the attribution of the SN~1987A-like
   events to SNe~IIP.
As noted earlier, a dome-like shape of the light curves of SN~1987A and SN~2000cb
   reflects different physics of the luminosity at the photospheric phase.
In the normal SNe~IIP the bulk of the radiated energy is an internal energy
   deposited during the shock wave propagation.
This is not true for the SN~1987A-like events because their dome-like light
   curves are powered by the radioactive decay, while the thermal energy
   is completely exhausted during the first 10 to 20 days.
We propose therefore to distinguish the SN~1987A-like events emphasizing
   their specific light curve shape by a notation ``SN~IId" where symbol ``d"
   stands for the ``dome".
This would reflect both the apparent observational property and the dominant
   role of the radioactivity in the light curve shaping.

%===============================================================================
\section{Conclusions}
\label{sec:concl}
%-------------------------------------------------------------------------------
Our goal was to derive the parameters of the peculiar type IIP SN~2000cb from
   the available observational data by employing hydrodynamical modeling and
   calculating hydrogen line profiles.
We conclude that SN~2000cb was the high-energy explosion of a massive BSG,
   which was more energetic than SN~1987A by a factor of three.
At the same time, these events are close counterparts by their pre-SN structure
   and by the common physics of their dome-like light curves.
SN~2000cb can thus be considered as a high-energy version of SN~1987A.

The progenitor mass, the explosion energy, and the radioactive $^{56}$Ni mass
   of SN~2000cb are consistent with the correlations revealed by the
   well-observed SNe~IIP studied hydrodynamically.
This combined with the clear evidence of one-sided $^{56}$Ni ejecta,
   which is rather ubiquitous among SNe~IIP, favors a common explosion mechanism
   for all these SNe.
We stress that the high explosion energy of SN~2000cb is far beyond the
   possibilities of the present-day versions of the neutrino-driven or
   magneto-rotational mechanism.

%===============================================================================
\begin{acknowledgements}
%-------------------------------------------------------------------------------
We thank Io K. W. Kleiser for sending us spectra of SN~2000cb.
V. P. U. is grateful to Wolfgang Hillebrandt for the possibility of working
   at the MPA, and his research has been supported in part by RFBR under grant
   10-02-00249-a.
%-------------------------------------------------------------------------------
\end{acknowledgements}
%-------------------------------------------------------------------------------

%===============================================================================

%-------------------------------------------------------------------------------

\begin{thebibliography}{}
%-------------------------------------------------------------------------------

\bibitem[2010]{ACJHH_10}
   Anderson, J. P., Covarrubias, R. A., James, P. A., Hamuy, M., \&
   Habergham, S. M. 2010, MNRAS, 407, 2660

\bibitem[2008]{BMA_08}
   Bisnovatyi-Kogan, G. S., Moiseenko, S. G., \& Ardeljan, N. V.
   2008, Astron. Reports, 52, 997

\bibitem[1981]{Che_81}
   Chevalier, R. A. 1981, Fundam. Cosmic Phys., 7, 1

\bibitem[1995]{CD_95}
   Chevalier, R. A., \& Dwarkadas, V. V. 1995, ApJ, 452, L45

\bibitem[1991]{Chu_91}
   Chugai, N. N. 1991, SvAL, 17, 400

\bibitem[2007]{Chu_07}
   Chugai, N. N. 2007, in AIP Conf. Proc. v. 937, Supernova 1987A: 20 years after.
   Supernovae and Gamma-ray Bursters, ed. S. Immler, K. Weiler, \&  R. McCray, 357

\bibitem[2000]{CU_00}
   Chugai, N. N., \& Utrobin, V. P. 2000, A\&A, 354, 557

\bibitem[2005]{DTMMN_05}
   Deng, J., Tominaga, N., Mazzali, P. A., Maeda, K., \& Nomoto, K.
   2005, ApJ, 624, 898

\bibitem[1977]{FA_77}
   Falk, S. W., \& Arnett, W. D. 1977, ApJS, 33, 515

\bibitem[1971]{GIN_71}
   Grassberg, E. K., Imshennik, V. S., \& Nadyozhin, D. K. 1971,
   Ap\&SS, 10, 28

\bibitem[1990]{HCE_90}
   Haas, M. R., Colgan, S. W. J., Erickson, E. F., et al. 1990, ApJ, 360, 257

\bibitem[1989]{HTS_89}
   Hanuschik, R. W., Thimm, G., \& Seidensticker, K. J. 1989, A\&A, 220, 153

\bibitem[2003]{HFWLH_03}
   Heger, A., Fryer, C. L., Woosley, S. E.,
   Langer, N., \& Hartmann, D. H. 2003, ApJ, 591, 288

\bibitem[1989]{HM_89}
   Hillebrandt, W., \& Meyer, F. 1989, A\&A, 219, L3

\bibitem[2004]{HMM_04}
   Hirschi, R., Meynet, G., \& Maeder, A. 2004, A\&A, 425, 649

\bibitem[1998]{IMN_98}
   Iwamoto, K., Mazzali, P. A., Nomoto, K., et al. 1998, Nature, 395, 672

\bibitem[2003]{KPSJM_03}
   Kifonidis, K., Plewa, T., Scheck, L., Janka, H.-Th., \& M\"uller, E. 2003,
   A\&A, 408, 621

\bibitem[2006]{KPSJM_06}
   Kifonidis, K., Plewa, T., Scheck, L., Janka, H.-Th., \& M\"uller, E. 2006,
   A\&A, 453, 661

\bibitem[2011]{KPK_11}
   Kleiser, Io K. W., Poznanski, D., Kasen, D., et al. 2011,
   MNRAS, 415, 372

\bibitem[1991]{MFA_91}
   M\"uller, E., Fryxell, B., \& Arnett, D. 1991, A\&A, 251, 505

\bibitem[2005]{PBB_05}
   Pastorello, A., Baron, E., Branch, D., et al. 2005,
   MNRAS, 360, 950

\bibitem[1989]{PJ_89}
   Podsiadlowski, Ph., \& Joss, P. C. 1989, Nature, 338, 401

\bibitem[2011]{RBNA_11}
   Rantsiou, E., Burrows, A., Nordhaus, J., \& Almgren, A.
   2011, ApJ, 732, 57

\bibitem[1988]{SNK_88}
   Saio, H., Nomoto, K., \& Kato, M. 1988, Nature, 334, 508

\bibitem[2008]{SJFK_08}
   Scheck, L., Janka, H.-Th., Foglizzo, T., \& Kifonidis, K. 2008,
   A\&A, 477, 931

\bibitem[2009]{Sma_09}
   Smartt, S. J. 2009, ARA\&A, 47, 63

\bibitem[2004]{Utr_04}
   Utrobin, V. P. 2004, Astron. Lett., 30, 293

\bibitem[2005]{Utr_05}
   Utrobin, V. P. 2005, Astron. Lett., 31, 806

\bibitem[2007]{Utr_07}
   Utrobin, V. P. 2007, A\&A, 461, 233

\bibitem[2005]{UC_05}
   Utrobin, V. P., \& Chugai, N. N. 2005, A\&A, 441, 271

\bibitem[2008]{UC_08}
   Utrobin, V. P., \& Chugai, N. N. 2008, A\&A, 491, 507

\bibitem[2009]{UC_09}
   Utrobin, V. P., \& Chugai, N. N. 2009, A\&A, 506, 829

\bibitem[2010]{UCB_10}
   Utrobin, V. P., Chugai, N. N., \&  Botticella, M. T. 2010, ApJ, 723, L89

\bibitem[1988]{WHT_88}
   Weiss, A., Hillebrandt, W., \& Truran, J. W. 1988, A\&A, 197, L11

\bibitem[1995]{WW_95}
   Woosley, S. E., \& Weaver, T. A. 1995, ApJS, 101, 181

\bibitem[1988]{WPE_88}
   Woosley, S. E., Pinto, P. A., \& Ensman, L. 1988, ApJ, 324, 466

\bibitem[1997]{WHWL_97}
   Woosley, S. E., Heger, A., Weaver, T. A., \& Langer, N. 1997
   [arXiv:astro-ph/9705146]

%-------------------------------------------------------------------------------
\end{thebibliography}
\end{document}